\documentclass[slac_one]{revtex4}
\usepackage{graphicx}
\usepackage{fancyhdr}
\usepackage{wrapfig,rotating,color}
\usepackage{amssymb,amsmath,array}
\begin{document}

% Title of paper
%%%%%%%%%%%%%%%%%%%%%%%%%%%%%%%%%%%%%%%%%%%%%%%%%%%%%%%%%%%%%%%%%%%%
\title{Precision Polarimetry at the International Linear Collider}
%-----------------------------------------
% \affiliation applies to all authors since the last \aff. command
\author{C.~Helebrant, D.~K\"afer, J.~List}
\affiliation{DESY, 22607 Hamburg, Germany}
%%%%%%%%%%%%%%%%%%%%%%%%%%%%%%%%%%%%%%%%%%%%%%%%%%%%%%%%%%%%%%%%%%%%
\begin{abstract}
  %%----------------------------------------------------------------------
  The International Linear Collider (ILC) will collide polarised electrons 
  and positrons at beam energies of 45.6 GeV to 250 GeV and optionally up 
  to 500 GeV. To fully exploit the physics potential of this machine, not 
  only the luminosity and beam energy have to be known precisely, but also 
  the polarisation of the particles has to be measured with an unprecedented 
  precision of $\Delta{\mathcal P} / {\mathcal P} = 0.25\%$ for both beams.

  An overall concept of high precision polarisation measurements at high 
  beam energies will be presented. The focus will be on the polarimeters 
  (up- and downstream of the $e^+e^-$ interaction point) embedded in the ILC 
  beam delivery system. Some challenges concerning the design of the Compton 
  spectrometers and the appropriate Cherenkov detectors for each polarimeter 
  are discussed. Detailed studies of photodetectors and their readout 
  electronics are presented focusing specifically on the linearity of the 
  device, since this is expected to be the limiting factor on the precision 
  of the polarisation measurement at the ILC.
  %%----------------------------------------------------------------------
\end{abstract}

\maketitle

\thispagestyle{fancy}

%============================================
\section{INTRODUCTION}\label{sec:intro}
%============================================
At the ILC, the overall scheme for a precise measurement of the beam polarisation 
conists in the usage of two Compton polarimeters per beam (up- and downstream of 
the $e^+e^-$ interaction point) and a polarisation measurement from annihilation 
data~\cite{bib:rdr, bib:epws-sum} to set the absolute scale.
 The two up- and downstream polarimeters are complementary, add redundancy and 
are used to intercalibrate the systems and reduce systematics. 
The aim is to reach a precision of $\Delta{\mathcal P}/{\mathcal P}=0.25\%$ 
with the polarimeters. If positron polarisation is realized, this can be 
improved to~$\Delta{\mathcal P}/{\mathcal P}=0.1\%$ by calibrating the 
absolute polarisation scale against annihilation data.

The high energy polarimeters are planned to consist of a spectrometer chicane 
and a Cherenkov detector to make use of the polarization dependence of Compton 
scattering, where the energies of the recoil electrons depend on the relative 
spin orientation of the original electrons and the laser photons. 
%---------------------------------------
On the order of $10^3$ beam electrons are scattered per laser interaction. 
Their energy distribution is then transformed into a spatial distribution by 
the magnetic chicane and measured by a multi-channel Cherenkov detector. 
A precise determination of the beam polarisation is achieved by measuring the 
asymmetry of the spectra obtained by switching the circular laser polarisation 
between +1 and -1. 
The system is designed such that the statistical uncertainty is negligible 
when averaging over more than $1$~s. Experience from previous polarimeters 
shows that the limiting systematic effect is the linearity of the detector 
response, especially the linearity of the photodetectors (PD) and readout 
electronics are of utmost importance~\cite{bib:phd-elia, bib:nonlin-eff}. 

%------------------------------
\begin{figure}[h!]
  \setlength{\unitlength}{1.0cm}
  \begin{picture}(16.0, 3.7)
    \put(-0.75, -0.20) {\includegraphics[width=0.60\textwidth]{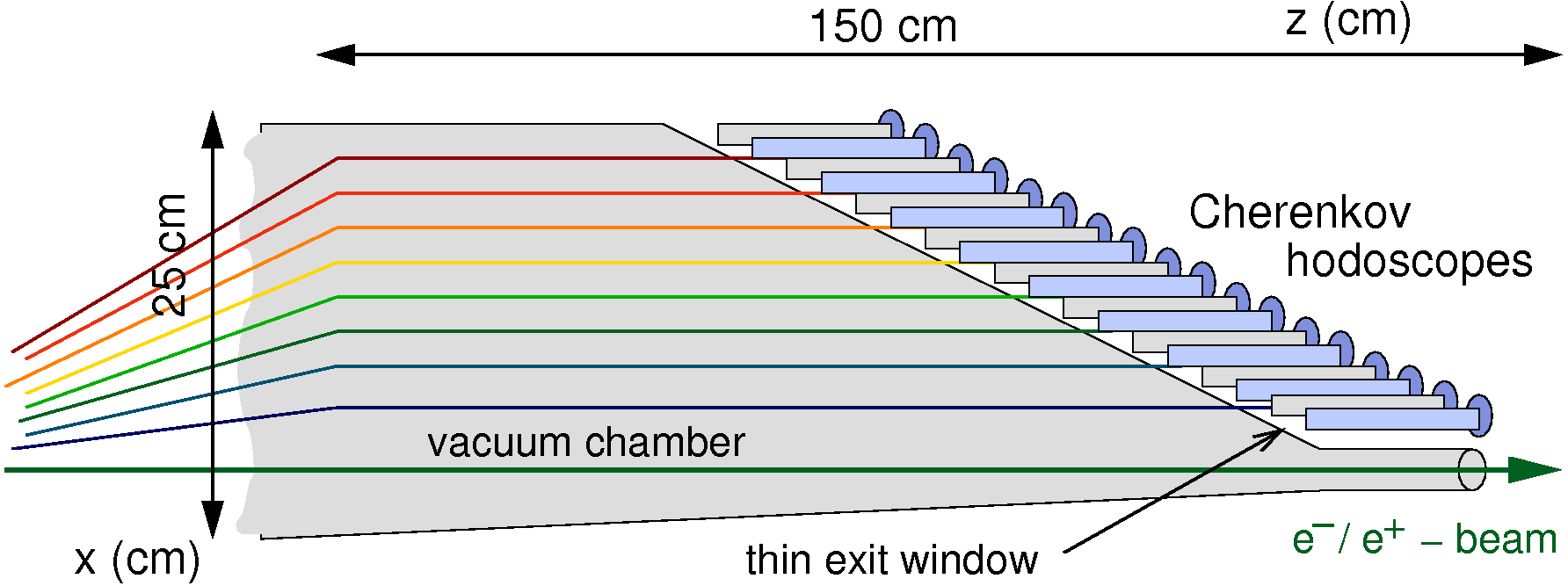}}
    \put(11.20, -0.15) {\includegraphics[width=0.32\textwidth]{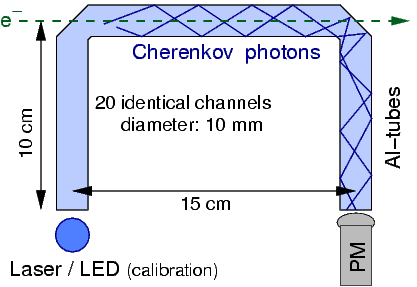}}
  \end{picture}
  \caption{\label{fig:cherdet}Schematic of the Cherenkov detector (left) 
    and one U-shaped, gas filled aluminum tube (right).\hspace*{4mm}}
\end{figure}
%------------------------------
\vspace*{-1.5mm} % under pic.distance
Several possibilities exist for the Cherenkov detector design 
(see Figure~\ref{fig:cherdet}). The first layout consists of staggered, U-shaped 
aluminum tubes filled with $C_4F_{10}$ as Cherenkov medium (high threshold of 10~MeV) 
and read out by conventional PDs. This technology is well-established, but the PDs 
are susceptible to magnetic fields and require high bias voltages.
A novel layout uses quartz fibers and silicon-based PDs (SiPM) to try and achieve 
a better spatial resolution and thus a more precise polarisation measurement. 
Semiconductor-based PDs promise to be more robust, less susceptible to magnetic 
fields and less expensive, but exhibit a lower Cherenkov threshold. One might, 
however, use both technologies in one system, adding complementarity and redundancy.

%===================================================================================
\section{PHOTODETECTOR  \& ELECTRONICS STUDIES}\label{sec:PDstudies}
%===================================================================================
A test facility has been set up to characterise different types of PDs regarding 
their adequacy for an ILC Cherenkov detector. It consists of a light-tight box, 
equipped with various mountings for different PDs, appropriate readout structures, 
the possibility to use optical filters and temperature sensors. 
A blue LED ($\lambda=470$ nm) connected to a function generator is used to 
illuminate the PDs and the data is read out via a high resolution double range 
12-bit VME charge-to-digital converter (QDC) with a $200$~fC and a $25$~fC wide 
least significant bit (LSB) per range. 
The aim is to measure and control the PD linearity to an order of $0.1\%$

%------------------------------
\begin{figure}[h!]
  \setlength{\unitlength}{1.0cm}
  \begin{picture}(16.0, 5.35)
    \put(-0.7, -0.30) {\includegraphics[width=0.47\textwidth]{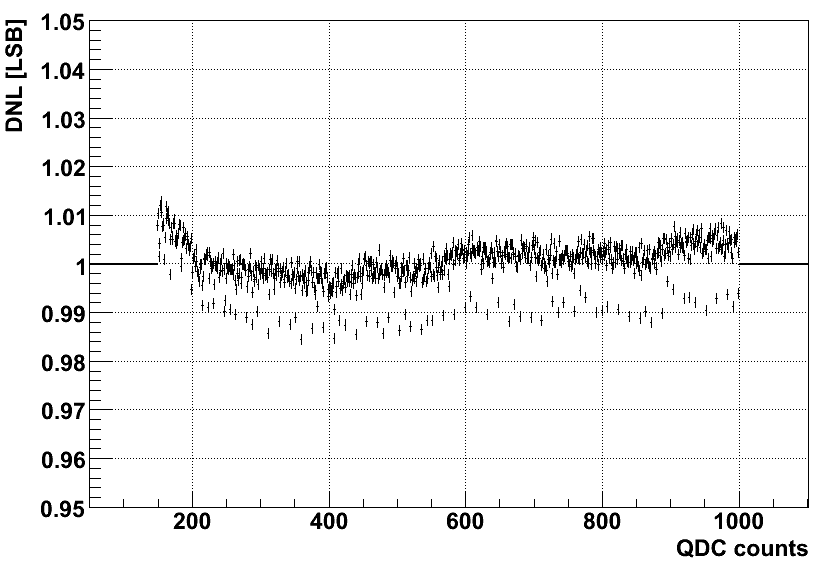}}
    \put( 8.4, -0.30) {\includegraphics[width=0.47\textwidth]{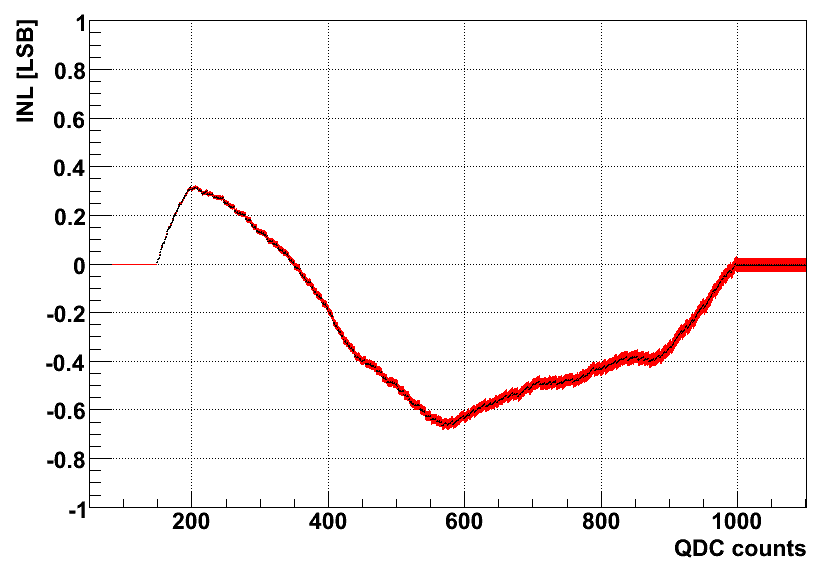}}
    %-------------------------------------------------------
    \put( 3.0,  4.85) {\sf from $10^9$ single measurements}
    \put(12.8,  4.85) {\sf maximum INL $= 0.64$ LSB}
  \end{picture}
  \caption{\label{fig:QDC-nonlin}Measurement of the QDC non-linearity: DNL (left) and resulting INL (right).}
\end{figure}
%------------------------------
\vspace*{-2mm} % under pic.distance
Since a considerable amount of non-linearity could be due to the electronics, 
the differential and integral non-linearity (DNL, INL) of the QDC have also been 
measured. A ramp of $f=10$~Hz is used as input signal, while a short random gate 
of $50$~ns is used to trigger the QDC. 
In case of an ideal QDC, one expects a uniform distribution of code bins. 
The ratios of measured and ideal distributions equal the code bin widths. 
These give the DNL, which is the deviation from the ideal code bin width 
of $1$~LSB. Summing the DNLs up to a specified bin gives the corresponding 
INL for this code bin. Figure~\ref{fig:QDC-nonlin} shows the results of a 
measurement with one billion single samples. The QDC shows good linearity 
over the entire range. The remaining non-linearity is within the range 
specified by the manufacturer and will be corrected for.

%------------------------------
\begin{wrapfigure}[13]{rH}{8.5cm}
  \setlength{\unitlength}{1.0cm}
  \begin{picture}(8.0, 5.2)
    \put( 0.0, 0.00) {\includegraphics[width=0.47\textwidth]{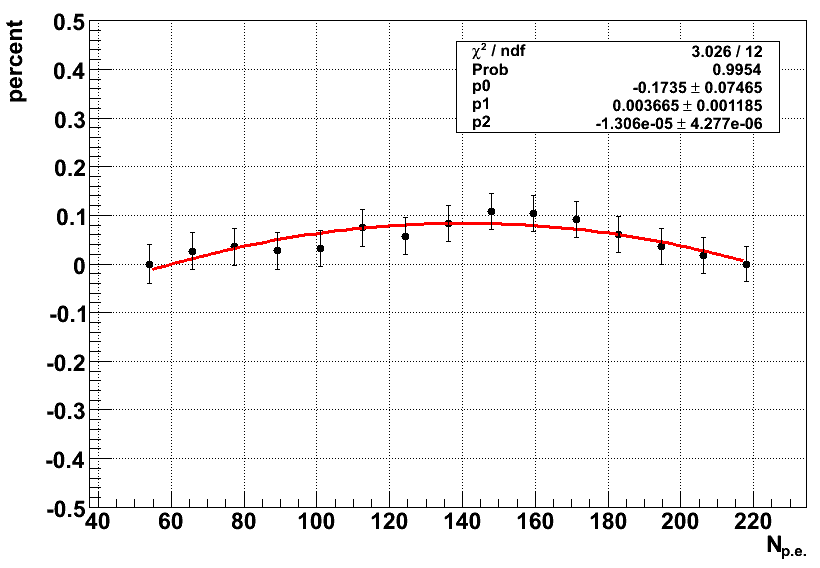}}
    \put( 2.3, 0.95) {\includegraphics[width=0.09\textwidth]{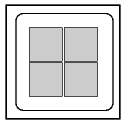}}
    \put( 4.0, 0.88) {\includegraphics[width=0.10\textwidth]{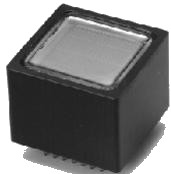}}
    %------------------------------------------------------
    \put( 6.1, 2.10) {\sf\scriptsize MAPM:}
    \put( 6.1, 1.65) {\sf\scriptsize anode structure}
    \put( 6.1, 1.20) {\sf\scriptsize and\quad photo}
  \end{picture}
  \vspace*{-4.5mm}
  \caption{\label{fig:PulseLength}Pulse-length method: PD non-linearity.}
\end{wrapfigure}
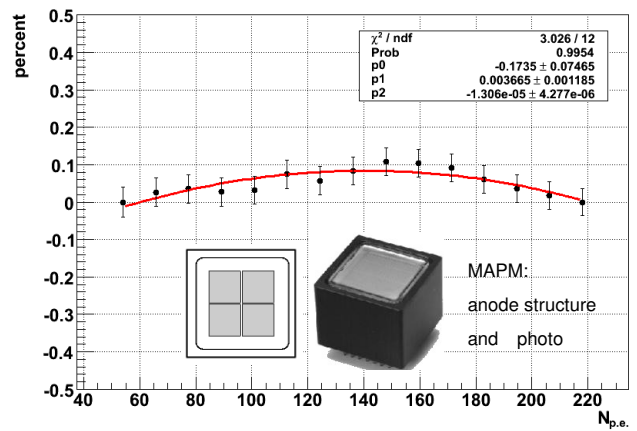
%------------------------------
A variety of PDs might be utilised in the Cherenkov detector, including 
conventional PMTs, 2$\times$2 multianode PMs (MAPM), and SiPMs. The following 
measurement, consisting of one million single events, has been made using a 
2$\times$2-MAPM (Hamamatsu R5900U-03-M4). 
A modified Poisson function is fitted to the QDC spectra to determine the 
number of incident photoelectrons. The reduced $\chi^2$ value of the fits 
is generally very good. 

Since the relation between light yield and bias voltage of the LED is not 
calibrated, several methods independent of the absolute scale of the LED have 
been established to measure the non-linearity of the photodetectors.
For the first method, the length of a rectangular pulse used to activate the LED 
is varied to ensure a linear variation of the amount of light on the photocathode.
The pulse length is varied in $5$~ns steps between $30$~ns and $100$~ns. As can 
be seen in Figure~\ref{fig:PulseLength}, the errors are very small ($\approx 0.05\%$). 
Thus, this method is suitable to observe a permille level non-linearity.

Two other methods sensitive to the DNL of the PDs have been developed and will 
soon be applied. One measures the difference in the PD's response when adding 
a second, very small LED pulse to the initial LED pulse. A second approach uses 
a four-holed mask applied to the PD, where the DNL is determined from the 
difference between the sum of single pulses through each hole separately 
and one pulse simultaneously through all four holes.
%---------------------------
Finally, optical filters can be employed to attenuate the light on the 
photocathode by defined amounts. Currently a set of filters is being 
calibrated to the neccessary precision.

%------------------------------
\begin{figure}[h!]
  \setlength{\unitlength}{1.0cm}
  \begin{picture}(16.0, 5.3)
    \put(-0.80, -0.30) {\includegraphics[width=0.48\textwidth]{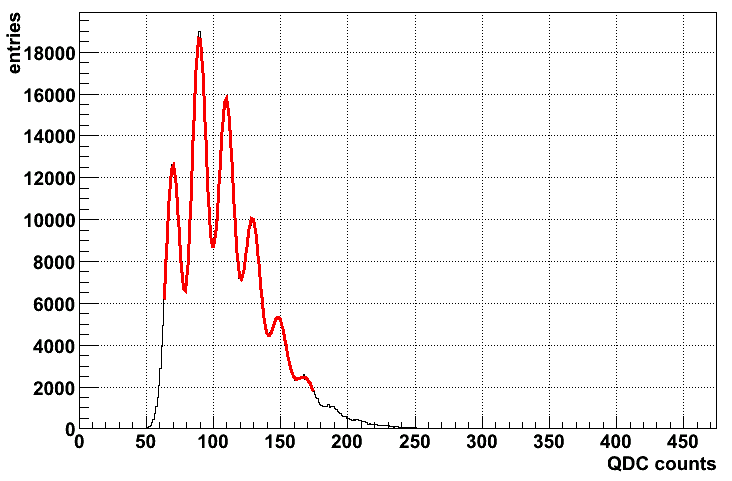}}
    \put( 4.82,  2.40) {\includegraphics[width=0.14\textwidth]{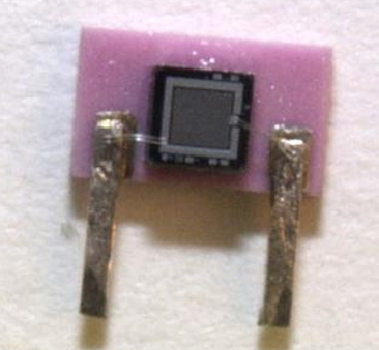}}
    \put( 8.30, -0.30) {\includegraphics[width=0.48\textwidth]{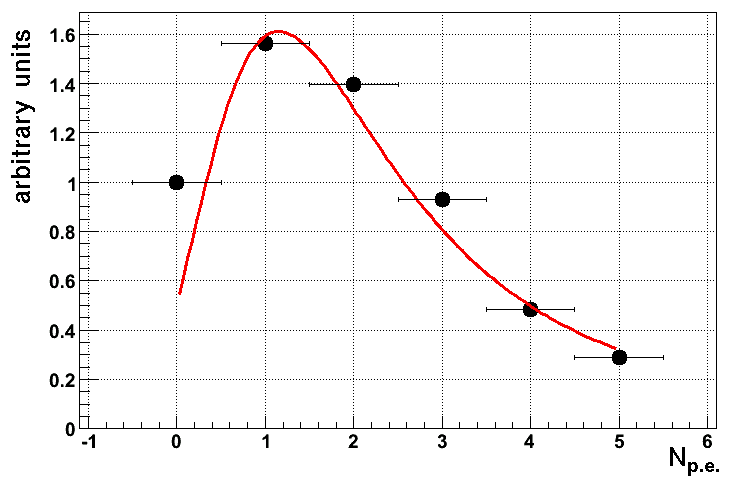}}
    %-------------------------------------------------------
    \put( 5.55,  4.74) {\sf SiPM photo}
  \end{picture}
  \caption{\label{fig:SiPM}SiPM (3$\times$3~mm$^2$) single photoelectron spectrum (left); 
    Landau fit to determine the MPV of $N_{p.e.}$ (right).}
\end{figure}
%------------------------------
\vspace*{-2mm} % under pic.distance
In addition, the pulse length method has been applied to a SiPM. A multi-gaussian 
fit to the single photoelectron spectrum is shown in Figure~\ref{fig:SiPM}~(left). 
The area under each peak is then plotted against the number of photoelectrons 
($N_{p.e.}$) and the resulting distribution is fitted with a gaussian-smeared 
Landau function to determine the most probable value (MPV) for the number of 
photoelectrons, see Figure~\ref{fig:SiPM}~(right). 
The analysis of this measurement is not yet completed and futher measurements 
whith different types of SiPMs are planned.

%=====================================================
\section{Conclusion and Outlook}\label{sec-concl}
%=====================================================
For the ILC the beam polarisation should be controlled to a yet unequaled
precision of $\Delta{\mathcal P}/{\mathcal P}=0.1\%$ by a combination 
of fast instrumental measurements and an absolute scale calibration from 
annihilation data.
Different methods to measure differential and integral linearity of 
photodetectors and electronics have been established and tested with 
a multianode photomultiplier. The pulse-length method is sensitive to 
non-linearities of the order of $0.1\%$. Once all methods have been 
checked thoroughly with the MAPM, further photodetector types will be 
analyzed and compared to each other.

% ****************************************************************************
% BIBLIOGRAPHY AREA
% ****************************************************************************

% ****************************************************************************
% END OF BIBLIOGRAPHY AREA
% ****************************************************************************

\end{document}